\documentclass[letter]{aa}
\usepackage{graphicx}
\usepackage[varg]{txfonts}
\begin{document}

\newcommand{\be}{\begin{equation}}
\newcommand{\ee}{\end{equation}}

\title{Primordial nucleosynthesis with varying fundamental constants}
\subtitle{Improved constraints and a possible solution to the Lithium problem}

\author{M. T. Clara\inst{1,2}
\and
C. J. A. P. Martins\inst{1,3}}
\institute{Centro de Astrof\'{\i}sica da Universidade do Porto, Rua das Estrelas, 4150-762 Porto, Portugal\\
\email{Carlos.Martins@astro.up.pt}
\and
Faculdade de Ci\^encias, Universidade do Porto, Rua do Campo Alegre, 4150-007 Porto, Portugal\\
\email{up201404911@fc.up.pt}
\and
Instituto de Astrof\'{\i}sica e Ci\^encias do Espa\c co, CAUP, Rua das Estrelas, 4150-762 Porto, Portugal}
\date{Submitted \today}

\abstract
{Primordial nucleosynthesis is an observational cornerstone of the Hot Big Bang model and a sensitive probe of physics beyond the standard model. Its success has been limited by the so-called Lithium problem, for which many solutions have been proposed. We report on a self-consistent perturbative analysis of the effects of variations in nature's fundamental constants, which are unavoidable in most extensions of the standard model, on primordial nucleosynthesis, focusing on a broad class of Grand Unified Theory models. A statistical comparison between theoretical predictions and observational measurements of ${}^4$He, D, ${}^3$He and, ${}^7$Li consistently yields a preferred value of the fine-structure constant $\alpha$ at the nucleosynthesis epoch that is larger than the current laboratory one. The level of statistical significance and the preferred extent of variation depend on model assumptions but the former can be more than four standard deviations, while the latter is always compatible with constraints at lower redshifts. If Lithium is not included in the analysis, the preference for a variation of $\alpha$ is not statistically significant. The abundance of ${}^3$He is relatively insensitive to such variations. Our analysis highlights a viable and physically motivated solution to the Lithium problem, which warrants further study.}


\titlerunning{Primordial Nucleosynthesis with Varying Fundamental Constants}
\authorrunning{Clara \& Martins}
\maketitle

\section{Introduction}
\label{introd}

Big Bang Nucleosynthesis (henceforth BBN) is among our best tools for probing the early universe. In the standard particle cosmology paradigm, it is highly predictive, yielding the abundances of all light nuclides as a function of one parameter (the baryon fraction) if it is assumed that all nuclear physics parameters are known. On the other hand, it can be used to place stringent constraints on a plethora of extensions to the standard model \citep{Steigman,Iocco,Pitrou}.

The unquestionable success of BBN is mitigated by the well-known Lithium problem \citep{Fields}. The theoretically expected and observed values differ by a factor of about 3.5 (the former being larger) which, given current uncertainties, represents a mismatch of more than five standard deviations. Although this is a long-standing problem, the solution is still unknown. It may involve systematics in astrophysical observations or in nuclear physics measurements of the required cross-sections, but it could also point to new physics beyond the standard paradigm. Recent discussions of the problem and some of its possible solutions can be found in \citet{Mathews} and in the BBN section of \citet{PDG}.

Nature is characterised by sets of physical laws and of fundamental dimensionless couplings; historically we have assumed that both are spacetime-invariant. The former assumption is manifestly a cornerstone of the scientific method but that is not the case for the latter: it is merely a simplifying assumption, albeit certainly a convenient one. We have no 'theory of constants' and their role in physical theories is not understood. Fundamental constants are simply defined operationally, just as any parameter whose value cannot be calculated within a given theory, but it must be found experimentally. Particle physics experiments show that fundamental couplings run with energy and in most extensions of the standard model, such as string theory \citep{Damour}, they are also unavoidably spacetime-dependent. Recent theoretical and observational overviews of the subject can be found in \citet{Uzan} and \citet{ROPP}.

The most actively pursued method to explore this scenario consists in experimentally or observationally testing the stability of the fine-structure constant, $\alpha=e^2/({\bar h}c)$, with astrophysical constraints typically expressed in terms of a relative variation with respect to the laboratory value, $(\Delta\alpha/\alpha)(z)=(\alpha(z)-\alpha_0)/\alpha_0$. Such a variation would have imprints in a wide range of physical mechanisms and environments \citep{Uzan} and increasingly tight constraints are emerging \citep{ROPP}. BBN is one such mechanism and the one probing the earliest cosmological epochs. The simplest approach in this context consists of allowing (phenomenologically) for a value of $\alpha$ that may differ from the present one while assuming that the rest of the physics is unchanged. This approach has been followed by several authors and leads to bounds on $\Delta\alpha/\alpha$ at the percent level \citep{Bergstrom,Pisanti,Nollett,Ichikawa}.

However, in most physically motivated extensions of the standard model, if $\alpha$ varies, it is expected that the other gauge and Yukawa couplings will also vary at some level \citep{Uzan,ROPP}. One exception is represented by the phenomenological models of \citet{Bekenstein}. Therefore, all such parameters should be permitted to vary, bearing in mind that in each specific model, these variations will be related to one another (the more parsimonious hypothesis being that they are all due to a single underlying physical mechanism) but these relations will themselves be model-dependent. This approach is expected to lead to more stringent constraints. Early efforts along these lines include the work of  \citet{Kawasaki}, who studied a simple particular model, as well as \citet{Muller}, who considered a more general theoretical framework but only analysed the impact on ${}^4$He, and \citet{Landau}, who used a simple semi-analytic approach. Other authors have specifically studied the impact of quark mass variations on BBN \citep{Resonance2,Resonance3}. Last but not least, it has been noticed that sensitivities to resonance parameters of some relevant nuclear reactions impact the Lithium problem \citep{Resonance4}.  This is an additional source of uncertainty which we do not explicitly address here.

Two approaches exist which, in principle, enable a fully self-consistent analysis of the problem. In \citet{Coc} (see also \cite{Luo}) it was noted that one can generically write the relative variations of other couplings as the product of some constant coefficients and the relative variation of $\alpha$ and with some reasonable simplifying assumptions \citep{Campbell},  only two such coefficients are needed. Most, but not all, of the necessary quantities were presented in \citet{Coc} who then  compared the observational data to the model for some specific choices of these coefficients. For example, they emphasised the important role of the Deuterium binding energy, but did not discuss the binding energies of the other nuclides. On the other hand, \citet{Stern1} (see also \citet{Stern2}) explicitly discussed the dependence of all key BBN parameters on the underlying physics parameters but having thus obtained a very generic description with a large number of free parameters, they only explored a few specific model choices.

In both approaches, the authors note that varying constants might provide a solution to the Lithium problem. The main argument for this is simple: in such models, one generally expects that the effects of the varying couplings will be larger for  heavier nuclides and it is therefore plausible that some amount of variation will reconcile the theoretical and observed  ${}^7$Li abundances without significantly impacting those of the lighter nuclides. Here we bring together the two approaches to carry out a self-consistent perturbative analysis of the effects of variations in nature's fundamental constants on BBN, applicable to a broad class of Grand Unified Theory models. Our results generalise previous approaches, but we do find agreement with the earlier works to the extent that direct comparisons are possible.

\section{Relating fundamental parameters to BBN quantities}
\label{guts}

The simplest self-consistent way to phenomenologically describe models which allow for simultaneous variations of several fundamental couplings is to relate the various changes to those of a particular dimensionless coupling, typically the fine-structure constant $\alpha$. These relations will be model-dependent. We follow \citet{Coc} in considering a broad class of grand unification models (in which unification is assumed to occur at some unspecified high energy), where the weak scale is determined by dimensional transmutation, and in further assuming that the relative variation of all the Yukawa couplings is the same. Finally, we assume that the variation of the couplings is driven by a dilaton-type scalar field \citep{Campbell}. These assumptions will naturally lead to relations between different relative variations.

In this case, the electron mass will vary as
\begin{equation}
\frac{\Delta m_e}{m_e}=\frac{1}{2}(1+S)~\frac{\Delta\alpha}{\alpha}\,,
\end{equation}
where $S$ is a dimensionless parameter related to electroweak physics (since fundamental particle masses are the product of the Higgs vacuum expectation value $\nu$ and the corresponding Yukawa coupling  $h$) defined as
\begin{equation}
\frac{\Delta \nu}{\nu} = S \frac{\Delta h}{h}\,;
\end{equation}
we recall that here we assume that all the Yukawa couplings vary in the same way. Since the proton is a composite object (for which the masses of the individual quarks actually have a relatively small contribution), modelling it introduces a second dimensionless parameter, $R$ related to Quantum Chromodynamics,
\begin{equation}
\frac{\Delta \Lambda_{QCD}}{\Lambda_{QCD}} = R \frac{\Delta \alpha}{\alpha}
\end{equation}
where $\Lambda_{QCD}$ is the QCD mass scale. This leads to a proton mass variation of
\begin{equation}
\frac{\Delta m_p}{m_p}=\big[0.8R+0.2(1+S) \big]~\frac{\Delta\alpha}{\alpha}\,.
\end{equation}
In this perturbative approach, the relative variations of the neutron mass (denoted $m_n$) and the average nucleon mass (denoted $m_N$) have the same behaviour,
\begin{equation}
\frac{\Delta m_n}{m_n}=\frac{\Delta m_N}{m_N}=\frac{\Delta m_p}{m_p}\,,
\end{equation}
while for Newton's constant we have
\begin{equation}
\frac{\Delta G_N}{G_N}=2\frac{\Delta m_p}{m_p}=\big[1.6R+0.4(1+S) \big]~\frac{\Delta\alpha}{\alpha}\,.
\end{equation}

The same treatment can be applied to other quantities relevant for BBN. The analysis of \citet{Coc} shows that the mass difference between neutrons and protons, $Q_N=m_n-m_p$,  changes as
\begin{equation}
\frac{\Delta Q_N}{Q_N}=[0.1+0.7~S-0.6~R]\, \frac{\Delta\alpha}{\alpha}\,,
\end{equation}
for the neutron lifetime, it is found that\begin{equation}
\frac{\Delta \tau_n}{\tau_n}=[-0.2-2.0~S+3.8~R]\, \frac{\Delta\alpha}{\alpha}\,;
\end{equation}
and for the deuterium binding energy
\begin{equation}
\frac{\Delta B_D}{B_D}=[-6.5(1+S)+18R]\, \frac{\Delta\alpha}{\alpha}\,.
\end{equation}
These authors do not explicitly study the behaviour of the binding energies of other nuclides but they do provide a relation between the relative variations in the nucleon and light quark masses,
\begin{equation}\label{quarkmass}
\frac{\Delta m_N}{m_N}\approx0.052 \frac{\Delta m_q}{m_q}\,,
\end{equation}
which is helpful for that purpose.

Here, $R$ and $S$ can be taken as free phenomenological parameters. Their absolute values can be anything from order unity to several hundreds, but while $R$ can be positive or negative (with the former case being more likely), but in terms of physical parameters, one expects that $S\ge0$. In any case, we can simply treat both as phenomenological parameters to be constrained by astrophysical data. In what follows, we consider two specific examples of these models, each described by particular values of $R$ and $S$. Firstly, current (possibly naive) expectations regarding unification scenarios suggest that typical values for the two parameters are \citep{Coc,Langacker}
\begin{equation}
R\sim36\,,\quad S\sim160\,;
\end{equation}
we refer to this as the Unification model. Although these numbers may be representative, they are certainly not unique. As an example, we take the dilaton-type model discussed by \citet{Nakashima}, which finds
\begin{equation}
R\sim109.4\,,\quad S\sim0\,;
\end{equation}
we refer to this as the Dilaton model. We also consider the general case where the parameters $R$ and $S$ are allowed to vary and are then marginalised. 

\section{Sensitivities of BBN observables}
\label{dent}

Generically speaking, the sensitivity of the primordial abundances on the various relevant particle physics parameters can be described as
\begin{equation}
\frac{\Delta Y_i}{Y_i}=\sum_j C_{ij}\frac{\Delta X_j}{X_j}\,,
\end{equation}
where $C_{ij}=\partial\ln{(Y_i)}/\partial\ln{(X_j)}$ are the sensitivity coefficients. Here (and in what follows) the perturbation is always done with respect to the standard model values. A detailed analysis can be found in \citet{Stern1} \citep[see also][]{Stern2,Pitrou} and we summarise the relevant sensitivity coefficients in Table \ref{table1}. This table does not include sensitivities to other cosmological parameters, such as the baryon fraction or the effective number of neutrinos since, in the present work, we assume that these have the standard values. We leave an extended analysis aside for future works.

For the parameters in the top half of the table, as well as for the Deuterium binding energy, the relations derived in the previous section can already be used to express the sensitivities as a function of the unification parameters $R$ and $S$ and the relative variation of $\alpha$. The same can be done for the other binding energies,whose sensitivities have been studied by \citet{Stern1} and by \citet{Resonance1}; in what follows we use as an intermediate step the following approximate relations which are also provided by \citet{Stern1}
\begin{equation}
\frac{\Delta B_T}{B_T}=-0.047\frac{\Delta\alpha}{\alpha}-2.1\frac{\Delta m_q}{m_q}\,,
\end{equation}
\begin{equation}
\frac{\Delta B_{^{3}He}}{B_{^{3}He}}=-0.093\frac{\Delta\alpha}{\alpha}-2.3\frac{\Delta m_q}{m_q}\,,
\end{equation}
\begin{equation}
\frac{\Delta B_{^{4}He}}{B_{^{4}He}}=-0.030\frac{\Delta\alpha}{\alpha}-0.94\frac{\Delta m_q}{m_q}\,,
\end{equation}
\begin{equation}
\frac{\Delta B_{^{7}Li}}{B_{^{7}Li}}=-0.046\frac{\Delta\alpha}{\alpha}-1.4\frac{\Delta m_q}{m_q}\,,
\end{equation}
\begin{equation}
\frac{\Delta B_{^{7}Be}}{B_{^{7}Be}}=-0.089\frac{\Delta\alpha}{\alpha}-1.4\frac{\Delta m_q}{m_q}\,,
\end{equation}
together with Equation (\ref{quarkmass}) from the previous section.

\begin{table}
\caption{Sensitivity coefficients of BBN nuclide abundances on the free parameters of our phenomenological parametrisation, defined in the main text.}
\label{table2}
\centering
\begin{tabular}{| c | c c c c |}
\hline
$C_{ij}$ & D & ${}^3$He & ${}^4$He & ${}^7$Li \\
\hline
$x_i$ & 42.0 & 1.27 & -4.6 & -166.6 \\
$y_i$ & 39.2 & 0.72 & -5.0 & -151.6 \\
$z_i$ & 36.6 & -89.5 & 14.6 & -200.9 \\
\hline
\end{tabular}
\end{table}

After some straightforward algebra, we can express the sensitivities of the BBN nuclides as
\begin{equation}
\frac{\Delta Y_i}{Y_i}=(x_i+y_iS+z_iR)\frac{\Delta\alpha}{\alpha}\,,
\end{equation}
where the sensitivity coefficients are listed in Table \ref{table2}. We note that the naive expectation that heavier nuclides should be more sensitive to variations is only approximately confirmed. The Lithium abundance is clearly the most sensitive one (which does suggest that a solution to the Lithium problem is possible in this context), but on the other hand, Helium-4 is less sensitive than Deuterium. Interestingly, Helium-3 is only sensitive to the $R$ parameter---an approximate cancellation means that its other two coefficients are comparatively very small. While we\ keep it in the analysis for completeness (and bearing in mind the caveats on the cosmological relevance of the current local measurements), we  find that it plays a minor role in the overall results.

\section{Constraints on $\alpha$ for specific and generic models}
\label{constraints}

We are now ready to compare this class of models with the observationally measured abundances. We  start by analysing some representative models (corresponding to specific choices of the parameters $R$ and $S$) and then move on to consider the general case where the parameters $R$ and $S$ are allowed to vary and are marginalised. For our analysis, we need to specify a fiducial standard model (yielding some theoretically predicted abundances), as well as the corresponding observationally measured values. For the former, we use the values published in the recent review by \citet{Pitrou}, while for the latter we rely on the recommended values in the 2017 Particle Data Group BBN review \citep{PDG} and its 2019 update (P. Molaro, private communication). All of these values are listed in Table \ref{table3}.

\begin{table}
\caption{Theoretical and observed primordial abundances used in our analysis. The theoretical ones have been obtained in \citet{Pitrou}. The observed ones are the recommended values in the 2019 Particle Data Group BBN review, with representative reference being, respectively, \citet{Aver}, \citet{Cooke}, \citet{Bania} and \citet{Sbordone}.}
\label{table3}
\centering
\begin{tabular}{c c c}
\hline
Abundance & Theoretical & Observed \\
\hline
$Y_p$ & $0.24709\pm0.00017$ & $0.245\pm0.003$ \\
$(D/H)\times 10^5$ & $2.459\pm0.036$ & $2.545\pm0.025$ \\
$({}^3He/H)\times 10^5$ & $1.074\pm0.026$ & $1.1\pm0.2$ \\
$({}^7Li/H)\times 10^{10}$ & $5.624\pm0.245$ & $1.6\pm0.3$ \\
\hline
\end{tabular}
\end{table}

We use a standard statistical likelihood analysis, with
\begin{equation}
\chi^2=\sum_i\frac{\left[Y_{i,obs}-Y_{i,th}\left(1+(x_i+y_iS+z_iR)(\Delta\alpha/\alpha)\right)\right]^2}{\sigma^2_{i,th}+\sigma^2_{i,obs}}
, \end{equation}
where the first terms in the numerator are the observed abundances and the second terms are the theoretically expected ones allowing for varying couplings. Theoretical and observational uncertainties are added in quadrature. The sum generically includes all four canonical abundances, but we  also report results obtained without ${}^3He$ (given the caveats in its observational determination) and without ${}^7Li$ (which can be seen as a null test of the sensitivity of our constraints since the theoretical and observational values of the remaining three abundances generally agree).

\subsection{Specific models}

\begin{figure*}[h!]
\centering
\includegraphics[width=8cm]{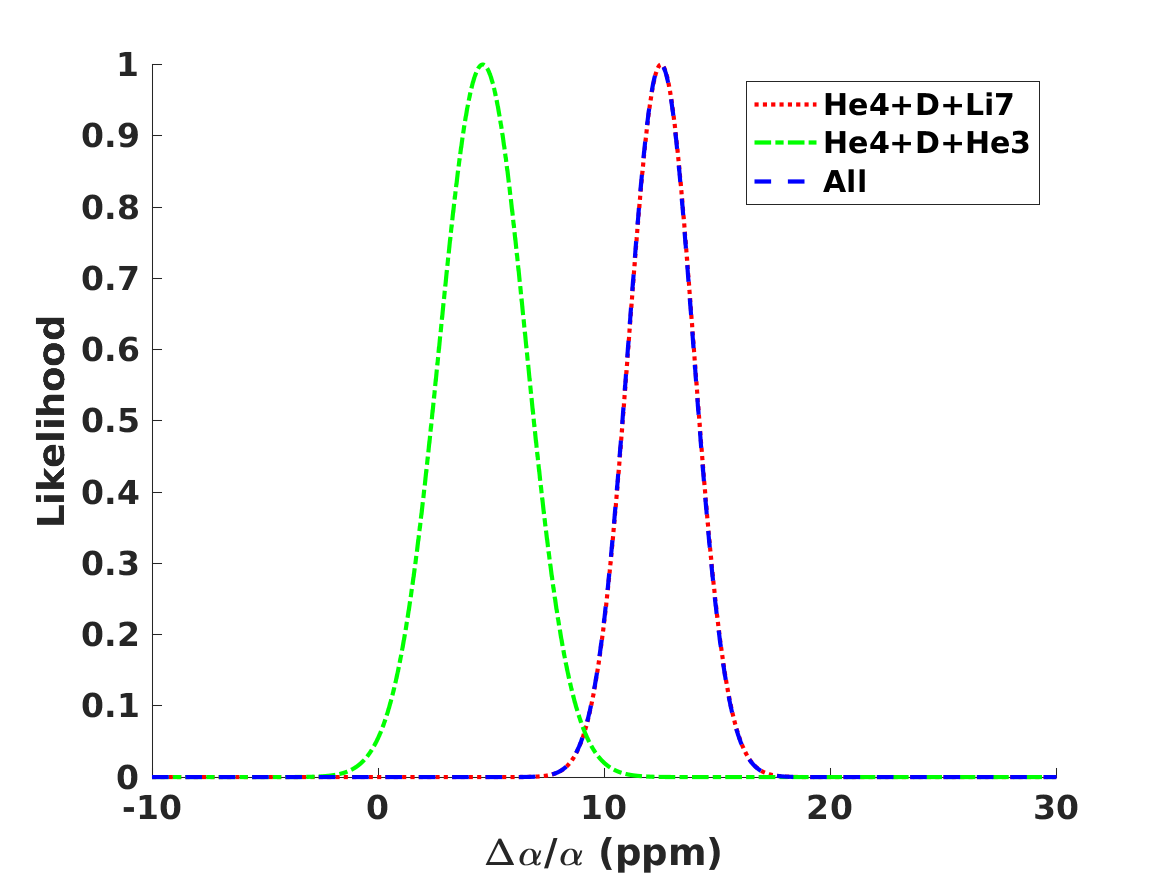}
\includegraphics[width=8cm]{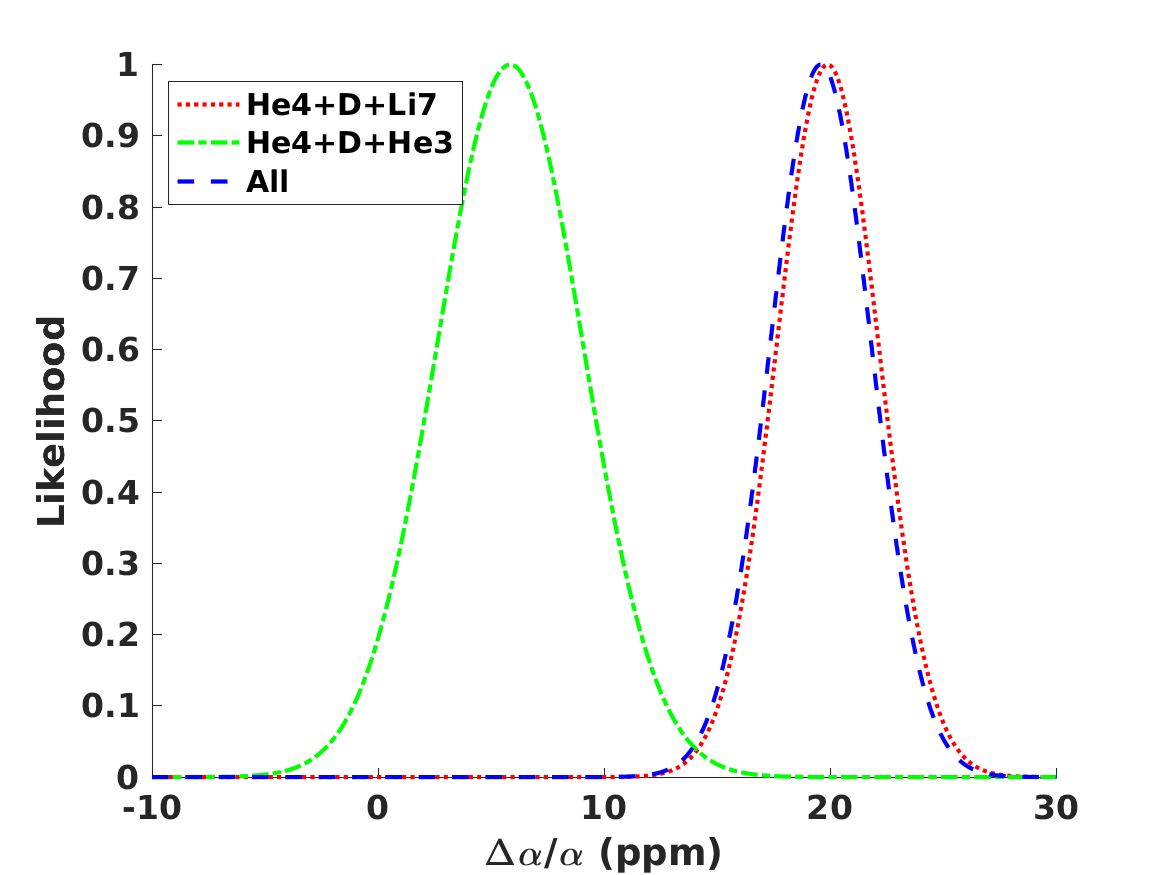}
\caption{Posterior likelihoods for the relative variation of $\alpha$ in the Unification and Dilaton models (left and right panels respectively), depending on the abundances considered in the analysis}
\label{figure1}
\end{figure*}

We start by constraining the three specific models already introduced in Section \ref{guts}. In these cases, the values of $R$ and $S$ are fixed, and the only free parameter is the value of $\Delta\alpha/\alpha$. The results of this analysis are depicted in Figure \ref{figure1} and summarised in Table \ref{table4}.

\begin{table}
\caption{Two-sigma ($95.4\%$ confidence level) constraints on $\Delta\alpha/\alpha$ for the two specific models considered according to the various choices of primordial abundances used in the analysis.}
\label{table4}
\centering
\begin{tabular}{c c c}
\hline
Abundances & Unification & Dilaton \\
\hline
${}^4$He+D+${}^7$Li & $12.5\pm2.9$ ppm & $19.9\pm4.5$ ppm \\
${}^4$He+D+${}^3$He & $4.6\pm3.8$ ppm & $5.8\pm6.5$ ppm \\
All four & $12.5\pm2.9$ ppm & $19.5\pm4.5$ ppm \\
\hline
\end{tabular}
\end{table}

Overall, we find that the data prefers positive values of $\Delta\alpha/\alpha$, corresponding to values of $\alpha$ at the BBN epoch that were larger then the present-day laboratory value. If  Lithium is not included in the analysis, this preference is not statistically significant (being at about the two-sigma level) but if Lithium is included a positive value is preferred at more than four standard deviations in each of the models. We also confirm the expectation that the Helium-3 measurement plays a relatively minor role: its impact is only visible for the Dilaton model, which is the one with the largest value of $R$ (recall that Helium-3 has almost no sensitivity to $S$).

\subsection{General case}

Since we are mainly interested in constraining $\Delta\alpha/\alpha$, we can, in principle, treat $R$ and $S$ as free parameters and then marginalise them. In practice there is one difficulty: given the lack on detailed knowledge of the physics of unification, it is difficult to identify a physically motivated choice of priors for both parameters. Remaining mindful of this fact, we agnostically consider three different assumptions as a way to assess the robustness of our constraints. Firstly, we take a narrow case, with uniform priors in the ranges $R=[-200,+200]$ and $S=[0,+400]$. Secondly, we take a broad case, with uniform priors in the ranges $R=[-500,+500]$ and $S=[0,+1000]$. Finally, we consider the broad case complemented by the prior constraint obtained in \citep{Clocks}, and coming from local laboratory experiments with atomic clocks \citep[for a more detailed discussion see][]{ROPP}
\begin{equation}
(1+S)-2.7R=-5\pm15\,;
\end{equation}
in what follows we refer to each of these as the Narrow, Broad, and Clocks scenarios respectively.

\begin{table}
\caption{Constraints on $\Delta\alpha/\alpha$ for the three scenarios considered, for various choices of primordial abundances used in the analysis. The listed values correspond to the best fit in each case and to the range of values within $\Delta\chi^2=4$ of it.}
\label{table5}
\centering
\begin{tabular}{c c c c}
\hline
Abundances & Narrow & Broad & Clocks \\
\hline
${}^4$He+D+${}^7$Li & $8.2^{+19.8}_{-4.2}$ ppm & $3.2^{+7.9}_{-1.7}$ ppm & $2.2^{+15.6}_{-0.6}$ ppm \\
${}^4$He+D+${}^3$He & $2.9^{+9.1}_{-2.7}$ ppm & $1.2^{+3.6}_{-1.2}$ ppm & $1.0^{+7.1}_{-0.9}$ ppm \\
All four & $7.4^{+18.8}_{-3.6}$ ppm & $3.0^{+7.6}_{-1.4}$ ppm & $2.2^{+15.5}_{-0.6}$ ppm \\
\hline
\end{tabular}
\end{table}

\begin{figure*}[h!]
\centering
\includegraphics[width=6cm]{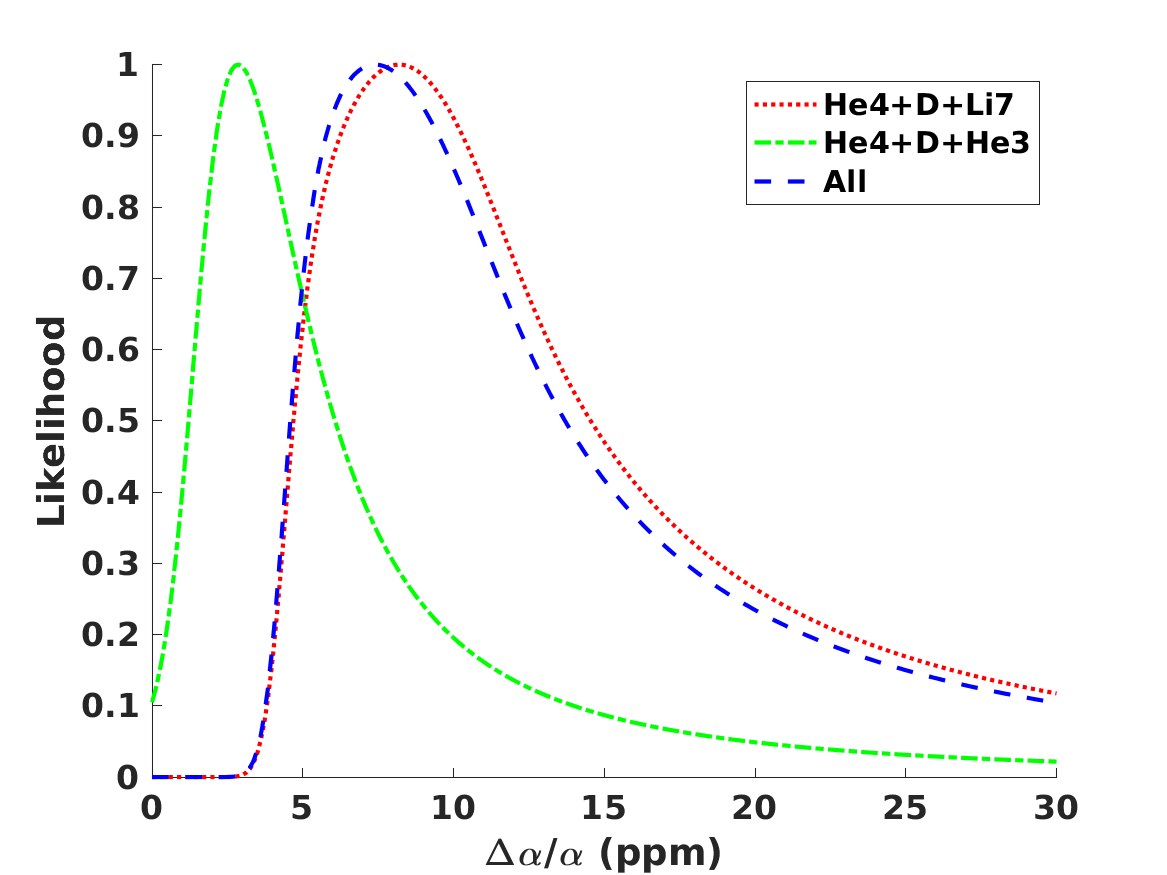}
\includegraphics[width=6cm]{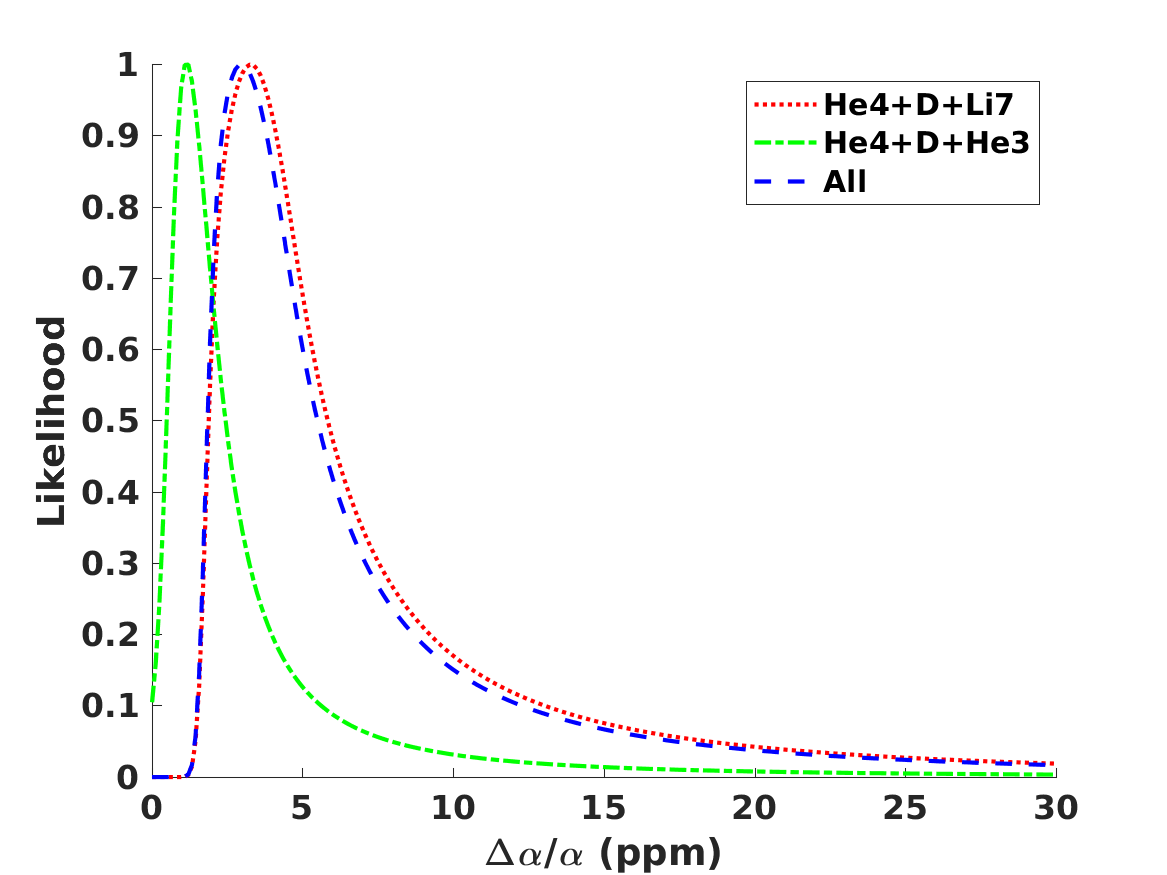}
\includegraphics[width=6cm]{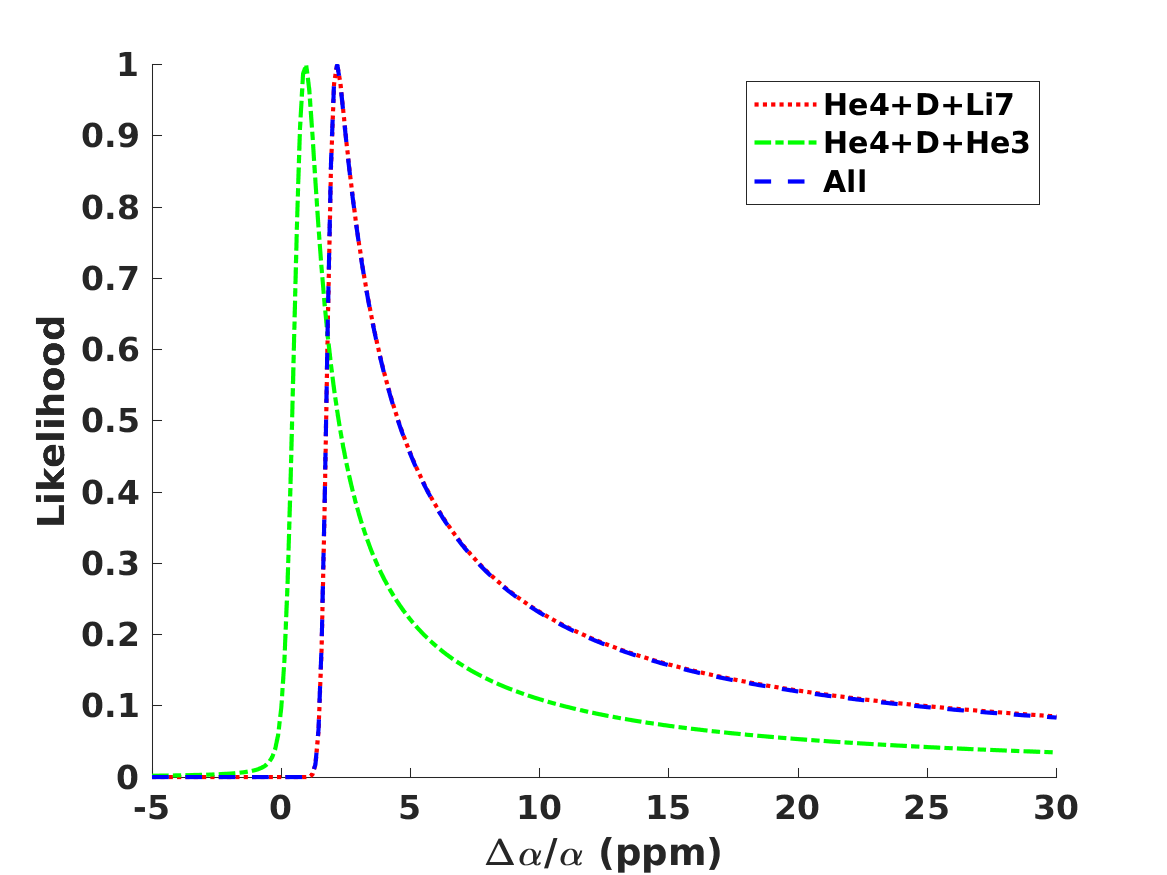}
\caption{Posterior likelihoods for the relative variation of $\alpha$ with $R$ and $S$ marginalised, in the Narrow, Broad and Clocks scenarios (from left to right), depending on the abundances considered in the analysis.}
\label{figure2}
\end{figure*}

The results are depicted in Figure \ref{figure2} and summarised in Table \ref{table5}. A first comment is that the posterior (marginalised) likelihoods for $\Delta\alpha/\alpha$ are manifestly non-Gaussian. For this reason, in Table \ref{table5} we report the best fit values in each case and the range of values within $\Delta\chi^2=4$ of it. The general trend highlighted in the previous section still holds: positive values of $\Delta\alpha/\alpha$ are clearly preferred, with the degree of statistical significance depending on whether or not the Lithium abundance is included.

There is also a significant dependence on the choice of priors. The Broad case leads to preferred values of $\alpha$ that are smaller (that is, closer to $\Delta\alpha/\alpha=0$) since a broader prior increases the fraction of the volume of parameter space with large values or $R$ and/or $S$. On the other hand, the atomic clocks constraint breaks the assumption of a uniform prior in the $R$--$S$ parameter space, preferring a particular slice thereof; the result of this is to bring best-fit values closer to the null result while further skewing the posterior likelihood towards values larger than the best-fit.

\section{Conclusions}
\label{concl}

We present a self-consistent perturbative analysis of the effects of variations in nature's fundamental constants on primordial nucleosynthesis. Such variations are unavoidable in most standard model extensions, and BBN is a powerful tool to constrain them. A statistical comparison of the latest theoretical and observed abundances yields a preferred value of $\alpha$ at the BBN epoch that is larger than the present one. The level of statistical significance of this preference depends on specific model assumptions: it is more than four standard deviations for the two representative models considered, less than that if  a phenomenological ensemble of all such models is considered. The preferred variations are at the ten parts per million level, which would be compatible will current constraints at lower redshifts, a summary of which can be found in \citet{ROPP}.

\begin{table*}
\caption{A comparison of the observed primordial abundances used in our analysis with our best-fit predictions (including both the baseline theoretical uncertainties and those on $\alpha$), in the Unification, Dilaton, and Clocks cases.}
\label{table6}
\centering
\begin{tabular}{c c c c c}
\hline
Abundance & Observed & Unification & Dilaton & Clocks \\
\hline
$Y_p$ & $0.245\pm0.003$ & $0.254\pm0.002$ & $0.244\pm0.001$ & $0.249\pm0.002$ \\
$(D/H)\times 10^5$ & $2.545\pm0.025$ & $2.68\pm0.07$ & $2.67\pm0.06$ & $2.58\pm0.06$ \\
$({}^3He/H)\times 10^5$ & $1.1\pm0.2$ & $0.88\pm0.05$ & $1.08\pm0.03$ & $1.01\pm0.04$ \\
$({}^7Li/H)\times 10^{10}$ & $1.6\pm0.3$ & $3.0\pm0.6$ & $3.8\pm0.5$ & $4.1\pm0.7$ \\
\hline
\end{tabular}
\end{table*}

\begin{figure*}[h!]
\centering
\includegraphics[width=8cm]{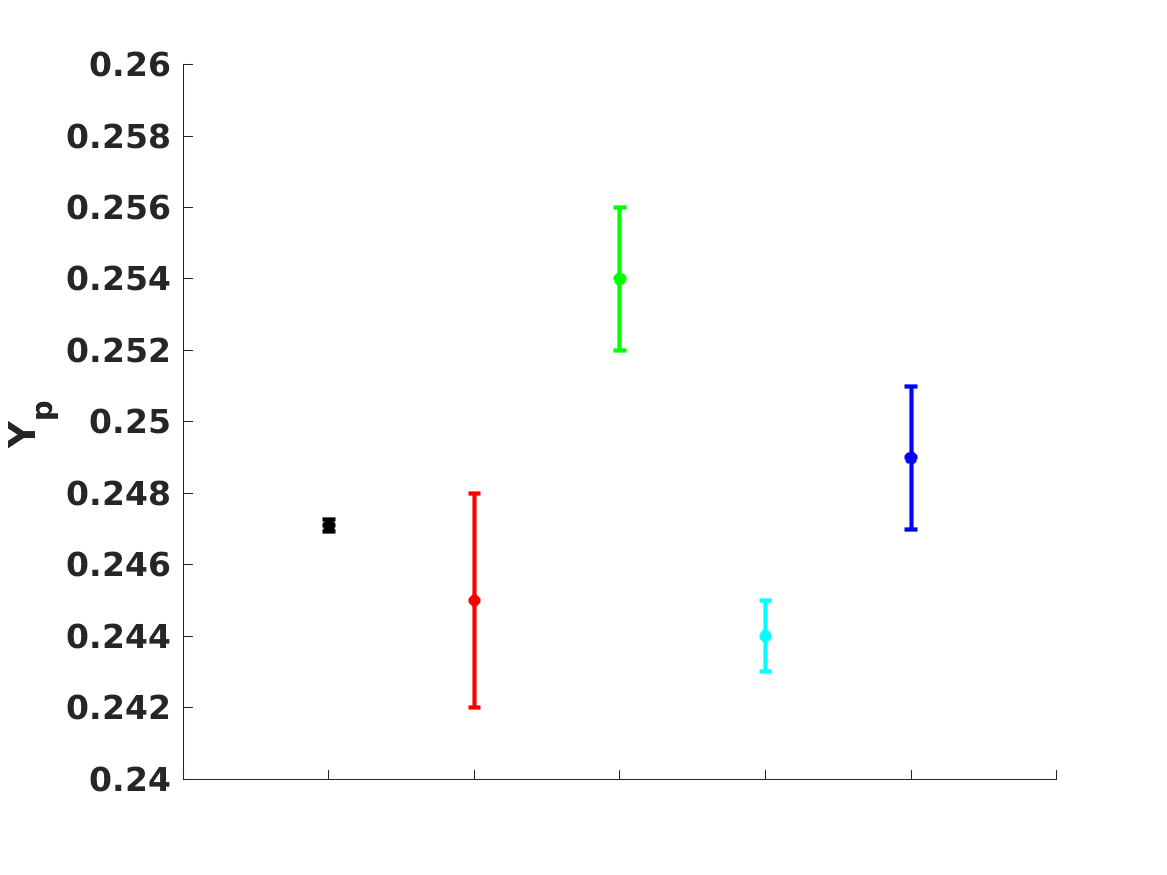}
\includegraphics[width=8cm]{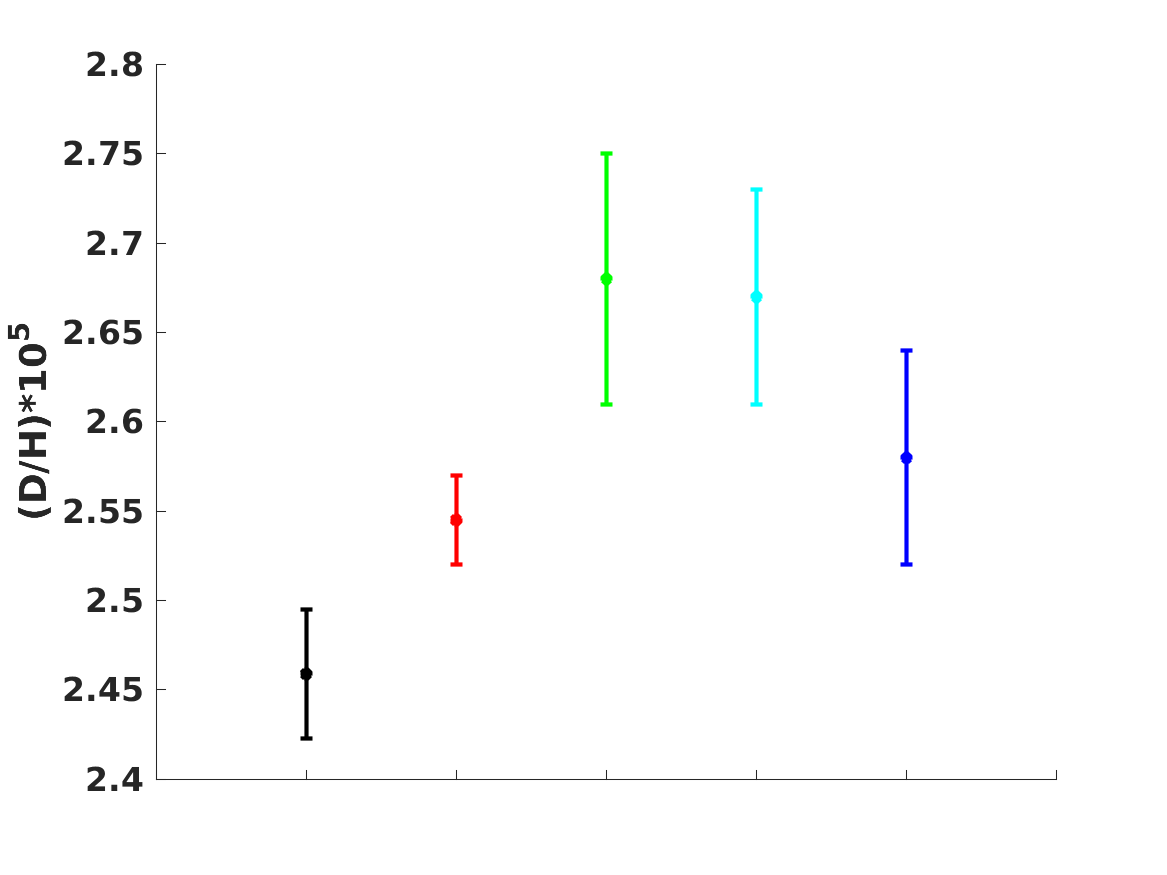}
\includegraphics[width=8cm]{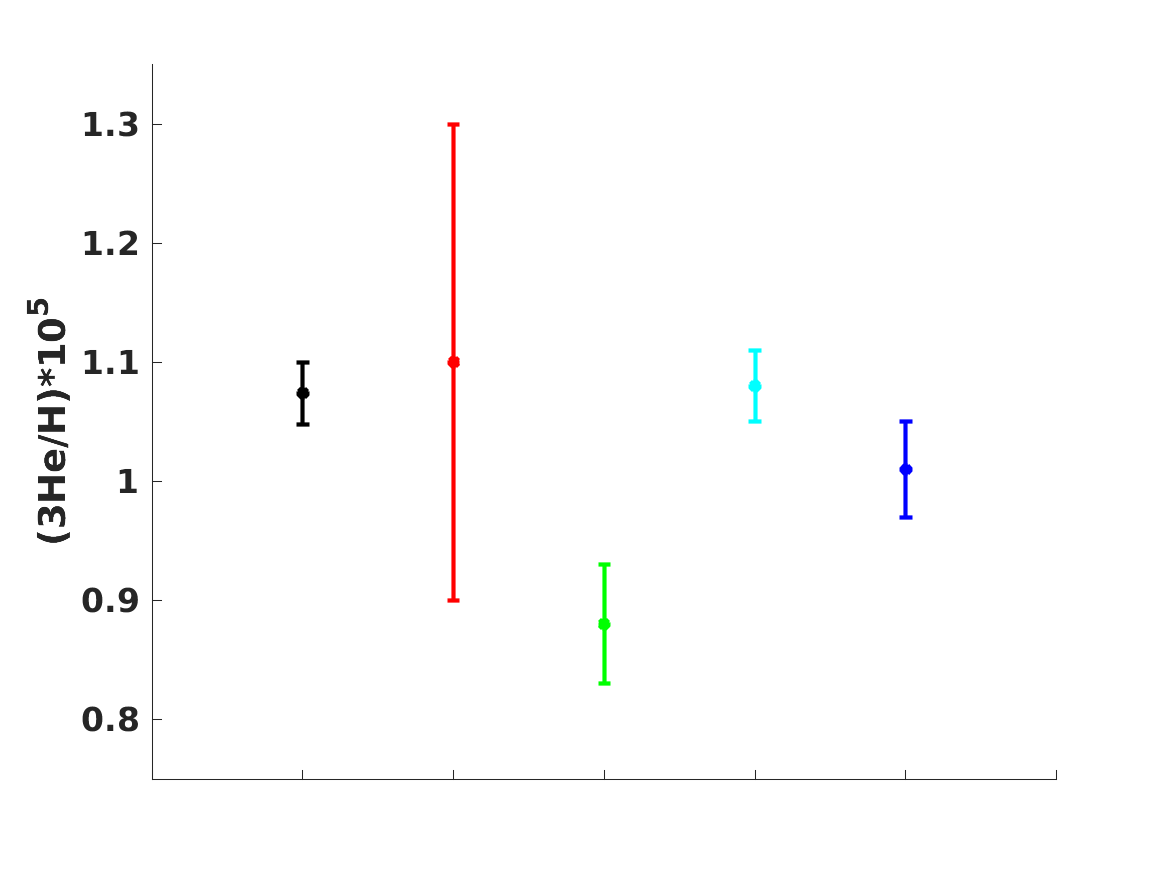}
\includegraphics[width=8cm]{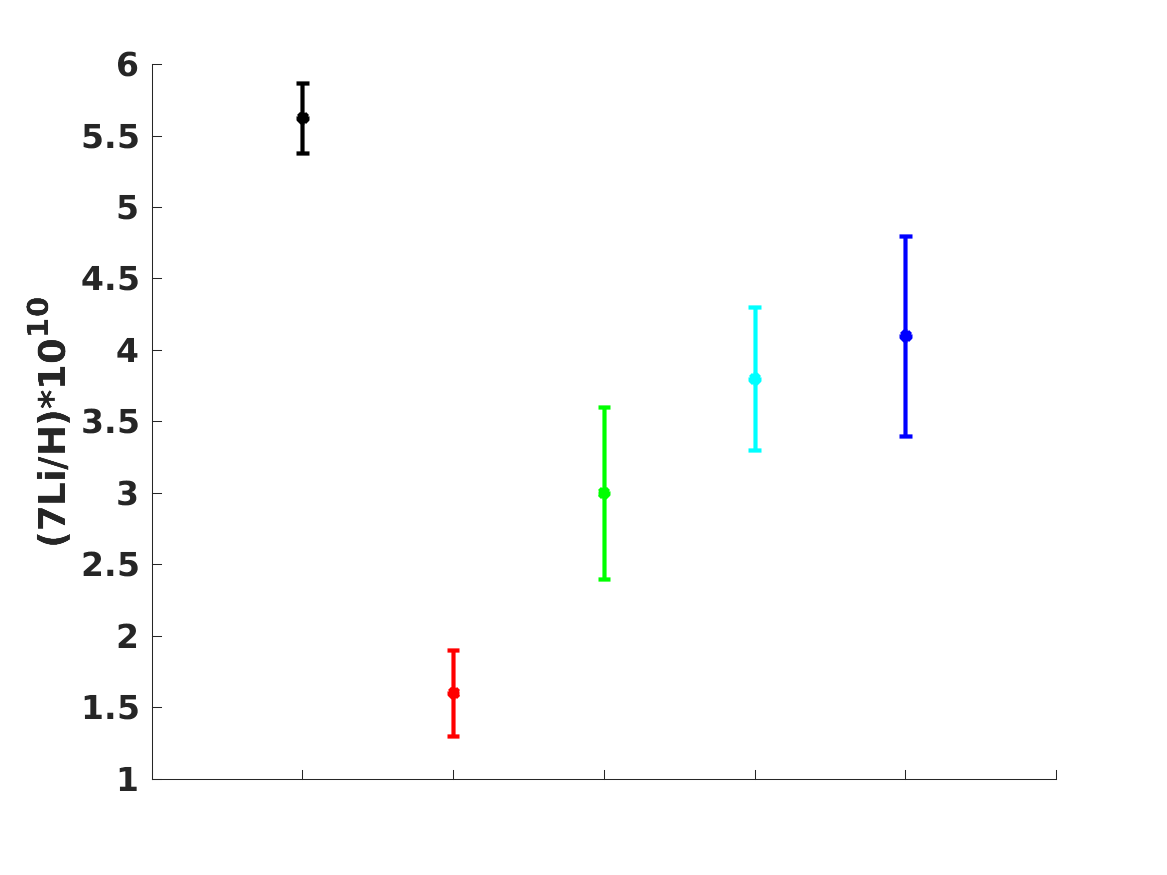}
\caption{Graphical comparison of the data in Tables \ref{table3} and \ref{table6} for each of the four nuclides. In each panel the black points are the standard theoretical values of \citet{Pitrou}, the red points are the observed abundances, and the green, cyan, and blue points are the expected values in the Unification, Dilaton, and Clocks cases considered in the present work. One sigma uncertainties have been depicted in all cases.}
\label{figure3}
\end{figure*}

We note that this preference is mainly (but not exclusively) driven by the Lithium abundance and we also find that Helium-3 is relatively insensitive to such variations. Table \ref{table6} compares the observed abundances with the theoretically expected values, including the best-fit values of $\alpha$ and their uncertainties, for the three main scenarios discussed in Section \ref{constraints}. All expected and observed values differ by no more than three standard deviations. A graphical comparison is in Figure \ref{figure3}. Our analysis, therefore, highlights a possible and physically motivated solution to the Lithium problem, confirming findings in previous works. Related issues which warrant additional study include degeneracies with cosmological parameters, such as the the baryon fraction and the effective number of relativistic species.


\begin{acknowledgements}
We are grateful to Paolo Molaro for useful correspondence, and to the anonymous referee for pertinent comments. This work was supported by FCT---Funda\c c\~ao para a Ci\^encia e a Tecnologia through national funds (PTDC/FIS-AST/28987/2017) and by FEDER---Fundo Europeu de Desenvolvimento Regional funds through the COMPETE 2020---Operacional Programme for Competitiveness and Internationalisation (POCI-01-0145-FEDER-028987). This work was supported by FCT/MCTES through national funds (PIDDAC) by grant UID/FIS/04434/2019.
\end{acknowledgements}

\bibliographystyle{aa} 
\bibliography{bbn} 

\begin{appendix}

\section{Additional tables and figures}

\begin{table}[h!]
\caption{Sensitivity coefficients of the BBN nuclide abundances on the relevant particle physics parameters, from \citet{Stern1} and \citet{Pitrou}. The parameters are, respectively, Newton's constant $G_N$, the fine-structure constant $\alpha$, the neutron lifetime $\tau_n$, the electron mass $m_e$, the mass difference between neutrons and protons, $Q_N=m_n-m_p$, the nucleon mass $m_N$ , and the binding energies of the various nuclides.}
\label{table1}
\centering
\begin{tabular}{| c | c c c c |}
\hline
$C_{ij}$ & D & ${}^3$He & ${}^4$He & ${}^7$Li \\
\hline
$G_{N}$ & 0.94 & 0.33 & 0.36 & -0.72\\
$\alpha$ & 2.30 & 0.79 & 0.00 & -8.10\\
$\tau_{n}$ & 0.422 & 0.141 & 0.732 & 0.438\\
$m_{e}$ & -0.16 & -0.02 & -0.71 & -0.82\\
$Q_{N}$ & 0.83 & 0.31 & 1.55 & 1.00\\
$m_{N}$ & 3.50 & 0.11 & -0.07 & -12.00\\
\hline
$B_{D}$ & -2.80 & -2.10 & 0.68 & 8.80\\
$B_{T}$ & -0.22 & -1.40 & 0.00 & -2.50\\
$B_{^{3}He}$ & -2.10 & 3.00 & 0.00 & -9.50\\
$B_{^{4}He}$ & -0.01 & -0.57 & 0.00 & -57.00\\
$B_{^{7}Li}$ & 0.00 & 0.00 & 0.00 & -6.90\\
$B_{^{7}Be}$ & 0.00 & 0.00 & 0.00 & 81.00\\
\hline
\end{tabular}
\end{table}

\end{appendix}
\end{document}